\documentclass[prl,aps,twocolumn,superscriptaddress,longbibliography]{revtex4-1}
\usepackage{latexsym}
\usepackage{lineno}
\usepackage{hyperref}
\usepackage{url}
\usepackage[caption=false,font={normalsize}]{subfig}
\usepackage{graphicx}
\usepackage{verbatim}
\usepackage{multirow}
\usepackage{amsmath, bm}
\newcommand{\beq}{\begin{eqnarray}}
\newcommand{\eeq}{\end{eqnarray}}
\usepackage{mathrsfs}
\usepackage{float}
\usepackage[usenames, dvipsnames]{color}
\usepackage{mathtools}
\usepackage{slashed}
\usepackage{physics}	
\usepackage{graphicx}   
\usepackage{epstopdf}
\usepackage{hyperref}   
\usepackage{bbold}
\usepackage{wasysym}
\usepackage{amssymb}
\usepackage{feynmp}
\usepackage[symbol]{footmisc}
\def \bs{\textbf}
\usepackage[latin1]{inputenc}
\usepackage{tikz}
\usetikzlibrary{decorations.pathmorphing}
\usetikzlibrary{shapes.misc}
\tikzset{cross/.style={cross out, draw=black, minimum size=8*(#1-\pgflinewidth), inner sep=0pt, outer sep=0pt},
cross/.default={1pt}}
\usetikzlibrary{patterns,math}
\newcommand{\RN}[1]{%
  \textup{\uppercase\expandafter{\romannumeral#1}}%
}

\newcommand{\qs}[1]{{\color{red} #1}}

\begin{document}
\title{
Shot noise 
and universal Fano factor as 
characterization of strongly correlated metals}
\author{Yiming Wang}
\affiliation{Department of Physics and Astronomy, Rice Center for Quantum Materials, Rice University, Houston, Texas 77005, USA}
\author{Chandan Setty}
\affiliation{Department of Physics and Astronomy, Rice Center for Quantum Materials, Rice University, Houston, Texas 77005, USA}
\affiliation{Department of Physics and Astronomy, Iowa State University, Ames, Iowa 50011, USA }
\affiliation{
Ames National Laboratory, U.S. Department of Energy, Ames, Iowa 50011, USA}
\author{Shouvik Sur}
\affiliation{Department of Physics and Astronomy, Rice Center for Quantum Materials, Rice University, Houston, Texas 77005, USA}
\author{Liyang Chen}
\affiliation{Department of Physics and Astronomy, Rice Center for Quantum Materials, Rice University, Houston, Texas 77005, USA}
\author{Silke Paschen}
\affiliation{Institute of Solid State Physics, Vienna University of Technology, 1040
Vienna, Austria}
\author{Douglas Natelson}
\affiliation{Department of Physics and Astronomy, Rice Center for Quantum Materials, Rice University, Houston, Texas 77005, USA}
\author{Qimiao Si}
\affiliation{Department of Physics and Astronomy, Rice Center for Quantum Materials, Rice University, Houston, Texas 77005, USA}

\begin{abstract}
Shot noise measures 
out-of-equilibrium current fluctuations and is a powerful tool to probe the 
nature of current-carrying excitations in quantum systems. Recent shot noise measurements in the 
heavy fermion strange metal YbRh$_2$Si$_2$ exhibit a strong suppression of the Fano factor ($F$) -- the ratio of the current noise to the average current in the DC limit.
This system is representative of metals
in which electron correlations 
are extremely strong.
Here we 
carry out the first theoretical study on 
the shot noise 
of 
diffusive metals in the regime of 
strong correlations.
A Boltzmann-Langevin equation 
formulation is constructed 
in a quasiparticle description 
in the presence of strong correlations.
We find 
that $F = \sqrt{ 3}/{4}$ 
in such a correlation regime.
{Thus, we establish the aforementioned Fano factor as universal to Fermi liquids, and show that }
the Fano factor suppression
observed in experiments on YbRh$_2$Si$_2$ 
necessitates a 
loss of the quasiparticles.
Our work opens the door to systematic theoretical studies of shot noise as a means of characterizing strongly correlated metallic phases and materials.
\end{abstract}

\maketitle
\paragraph*{{\bf Introduction: }}
It is standard to describe 
metallic systems with electron correlations 
in terms of quasiparticles. These are
elementary excitations
that carry the quantum numbers of a bare electron,
including charge $e$.
In strange metals near quantum 
criticality \cite{Kei17.1,Pas21.1}, however,
the current carriers
are expected to lose \cite{Hu-qcm2022,Phillips-science22} 
a well-defined quasiparticle 
interpretation 
and hence the notion of a discrete charge.
This issue is especially pronounced in quantum critical heavy fermion metals \cite{Pas21.1,KirchnerRMP,Coleman-Nature,StewartRMP}, for which 
a beyond-Landau description involving 
Kondo destruction 
\cite{Si-Nature,Colemanetal,senthil2004a} 
has received considerable experimental support
\cite{paschen2004,Friedemann.10,shishido2005,Schroder,Aro95.1,Prochaska2020}.\par
How to directly prove that quasiparticles are lost in correlated metals is largely an open question.
One established means of such a characterization is 
in terms of the ratio of the thermal and electrical conductivities ($\kappa$ and $\sigma$, respectively). The quasiparticle description requires
that 
the Lorenz number, $L \equiv \kappa /T \sigma $,
obeys the Wiedemann-Franz law \cite{chester1961,castellani1987}. Given that charge-neutral excitations such as phonons also contribute to the thermal current, 
alternative means of characterizing the absence of quasiparticles 
are much called for.
Here we 
address this issue in terms of shot noise~\cite{Buttiker2000} -- 
the out-of-equilibrium fluctuations
of the electrical current.

When electron correlations are strong, 
the shot noise of diffusive metals has not been theoretically considered. Here we
show that, 
with a suitable requirement on a hieararchy of 
length scales,
the shot noise 
Fano factor
($F$), defined as the ratio of average current fluctuations in the static limit 
to the average current, 
{has a univeral value $F=\sqrt{3}/4$ in Fermi liquids in the presence of Landau parameters and quasiparticle weight reduction. By extension, 
the quasiparticle description fails 
for strongly correlated metals
when their Fano factor}
disobeys $F=\sqrt{3}/4$.

Shot noise has 
proven 
invaluable in understanding several correlated electronic systems and materials such as quantum hall liquids, superconductors and quantum dots, and has offered a window into the nature of elementary charge carriers in their respective ground states~\cite{Hashisaka2021}. For example, measurement of the shot noise Fano factor
has been pivotal in uncovering the $\frac{1}{3}$ charge fractionalization in the Laughlin state~\cite{Mahalu1998, Etienne1997} and charge 2$e$ Cooper pairs in (fluctuating) superconductors~\cite{Natelson2019, Allan2021}.
Furthermore, the Fano factor has been widely used to isolate dominant scattering mechanisms in mesoscopic systems.
Typically, in a diffusive Fermi gas, $F= \frac{1}{3}$ when scattering is dominated by impurities~\cite{Buttiker1992, Nagaev1992}.
In a similar Fermi-gas-based approach, when the inelastic electron-electron scattering rate is allowed for and higher than the elastic scattering rate, but in the absence of significant electron-phonon scattering, 
 $F$ was shown to equal $\frac{\sqrt{3}}{4}$~\cite{Rudin1995, Nagaev1995}.  \par
Recently, shot noise measurements in mesoscopic wires of the heavy 
fermion compound YbRh$_2$Si$_2$ showed
a large suppression of the Fano factor well below that of a Fermi-gas-based 
diffusive metal~\cite{Natelson2022}. This was found to occur at low enough temperatures ($<10$K) where equilibration of electrons via a bosonic bath such as phonons is minimal, and cannot account for the reduced shot noise signal. Since the only other source of inelastic scattering is strong electron interactions, a natural question arises: can strong correlation effects in a 
Fermi liquid (FL) account for the 
observed shot noise suppression, 
or is it a signature of strange metallicity near the  quantum critical point? This issue is particularly important, given that, even for the FL state of heavy fermion metals, the effect of interactions is pronounced and can induce orders-of-magnitude renormalization of both the quasiparticle weight and effective interactions. 
\par
Here we report on the first theoretical study 
about 
the shot noise 
of strongly correlated 
diffusive metals.
We show that, in the presence 
of strong correlations,
the Fano factor of a diffusive FL 
$F$ is equal to $\frac{\sqrt 3}{4}$.
In particular, it is 
independent of quasiparticle weight or Landau FL parameters.
It includes, for example, the case when the quasiparticle residue $z_\bs k$ approaches an infinitesimally small (but nonzero) value, or when the Landau parameters are as large as in the heavy fermion metals (typically about $\sim 10^3$).
Since the existence of quasiparticles is central to the robustness of the Fano factor prediction, the 
experimentally
observed suppression~\cite{Natelson2022}
strongly indicates a loss of quasiparticles.
\par To this end, we derive the Boltzmann-Langevin transport equation for a diffusive metal in a regime suitable for addressing strong correlations. This allows us to analyze the role of strong interactions on the Fano factor.  First, we notice that charge conservation constrains the Boltzmann equation to be independent of the quasiparticle weight. This holds even in the presence of anisotropic effects when $z_{\bs k}$ is strongly momentum dependent. As a consequence, the current noise, average current, and Fano factor determined from the Boltzmann-Langevin equation are independent of $z_{\bs k}$. Second, we calculate the shot noise,
and demonstrate that the shot noise and average current get renormalized identically by the Landau parameters. As a result, the Fano factor remains robust to the introduction of arbitrarily strong interactions within FL theory. Also inherent in this cancellation is our observation that the  conductance entering both shot noise and average current are equal  and determined by the same quasiparticle lifetime, i.e., there is a symmetry of the scattering rate between single and multi-particle operators. This symmetry exists because in FLs, the interactions are instantaneous (no frequency dependence) and the electron scattering processes are Poissonian, i.e., independent of one another.
\par
\begin{figure}[!t]
\centering
\subfloat[\label{fig:1}]{%
  \includegraphics[width=0.575\columnwidth]{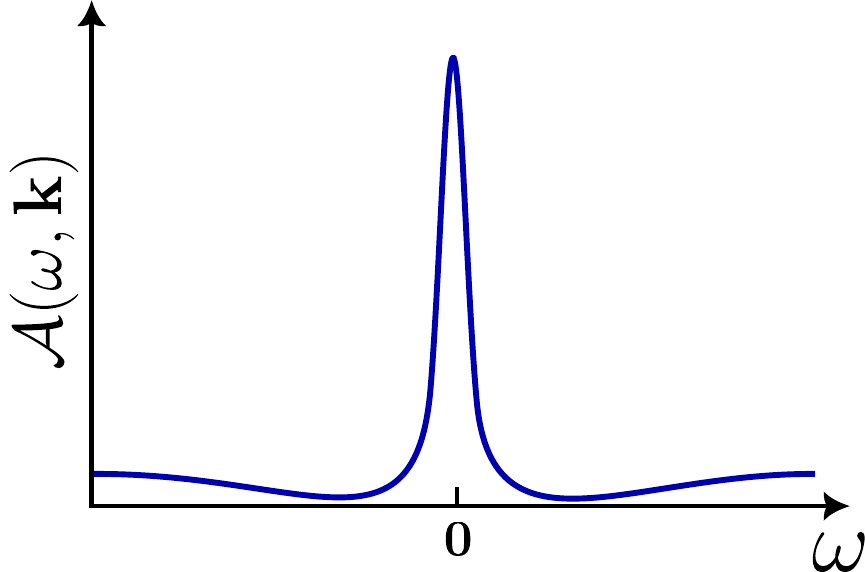}%
}
\hfill
\subfloat[\label{fig:2}]{%
  \includegraphics[width=0.3\columnwidth]{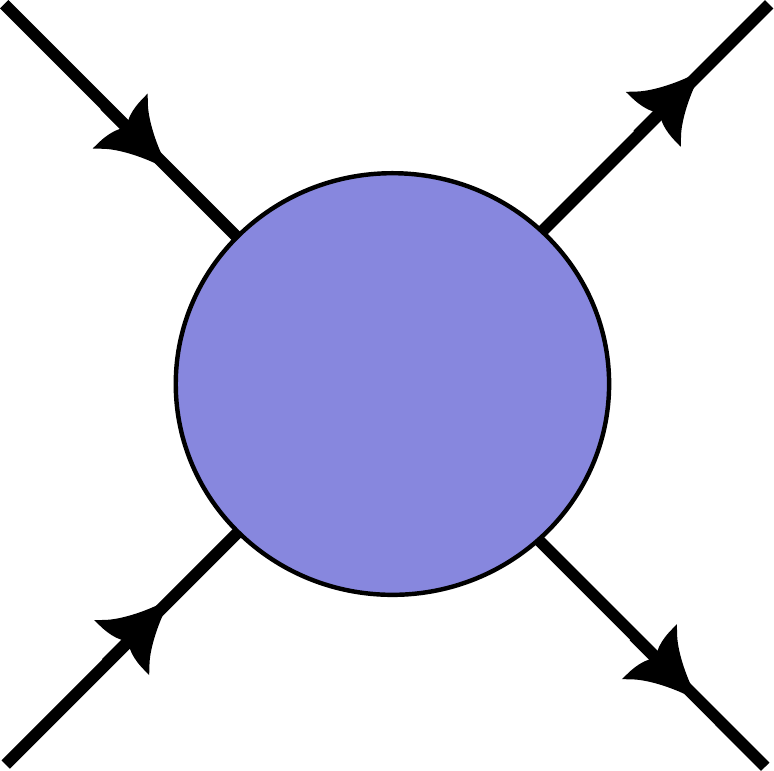}%
}
\caption{(a) Plot of the spectral function $A(\omega, \bs k)$ with $\bs k$ located at the Fermi momentum $\bs k_F$. The area of the peak is determined by $z_{\bs k}$. (b) A representation of the irreducible vertex contributing to the total vertex in the Bethe-Salpeter equation in Fig.\,\ref{Fig:BSE}(b)
.} \label{Fig:Spectral}
\end{figure}
In the remainder of the paper, we
construct the Boltzmann-Langevin equation 
in the strongly correlated regime of a FL.
We closely follow the diagrammatic analysis of Betbeder-Matibet and Nozieres~\cite{Nozieres1966}.  
In particular, we analyze the role of interactions via coupling through the scalar and vector potentials. 
We then calculate the Fano factor through the shot noise and average current,  and discuss its relevance to experiments in YbRh$_2$Si$_2$ before presenting our conclusions.

\paragraph*{{\bf Boltzmann-Langevin  equation for interacting Fermi liquids: }}
Interacting electrons in the presence of randomly distributed impurities is governed by the Hamiltonian,  $H = H_0 + H_I + H_{imp}$.
Here $H_0 = \sum_{\bs k, \sigma} \epsilon_{\bs k} c_{\bs k \sigma}^{\dagger} c_{\bs k \sigma}$, $H_I = \frac{1}{2}\sum_{\substack{\bs k, \bs k', \bs q \\ \sigma \sigma'}}  V(\bs q)c_{\bs k + \bs q \sigma}^{\dagger} c_{\bs k' -\bs q \sigma'}^{\dagger} c_{\bs k' \sigma'} c_{\bs k \sigma} $ describe the  non-interacting electron dispersion and Coulomb interaction,  respectively. In addition,
$H_{imp} = \sum_{i, \bs k, \bs q, \sigma} U(\bs q) e^{-i \bs q\cdot \bs R_i} c^{\dagger}_{\bs k +\bs q \sigma} c_{\bs k \sigma}$ marks electron scattering from dilute impurities.
The bare electron operator is denoted by $c_{\mathbf k \sigma}$ with $\mathbf k$ and $\sigma$ representing momentum and spin.  
The Coulomb and impurity matrix elements are denoted by $V(\bs q)$ and $U(\bs q)$, respectively.
Next we consider an external field with Fourier components $\lambda(\bs k)$ that couples to the interacting electrons, which contributes 
\beq
H'(\mathbf q_0, \omega_0) = \sum_{\bs k, \sigma} \lambda_\sigma(\bs k) c^{\dagger}_{\bs k + \bs q_0/2 \sigma}  c_{\bs k - \bs q_0/2 \sigma} e^{-i \omega_0 t}+ \mbox{h.c.},
\eeq
to the total Hamiltonian, where $(\omega_0, \bs q_0) \equiv q_0$ is the energy and momentum transfer between the electrons and the external field. 

\begin{figure}[tb]
\subfloat[\label{fig:del-n}]{%
  \includegraphics[width=\columnwidth]{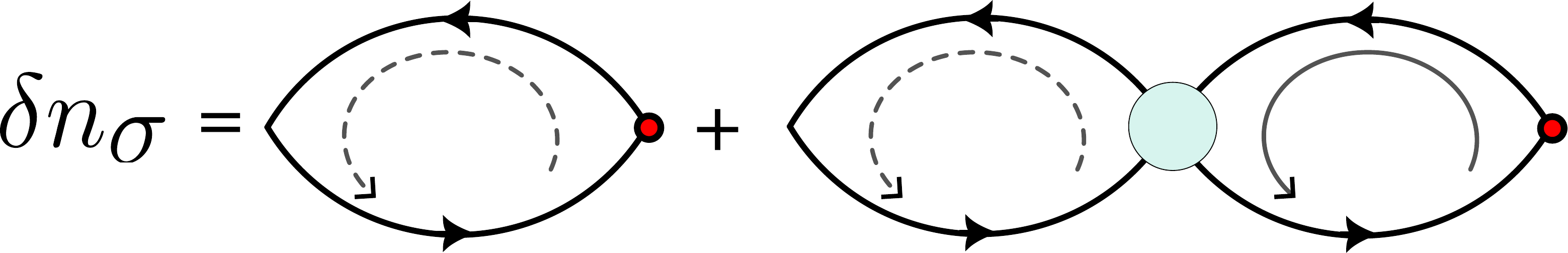}%
}
\hfill
\subfloat[\label{fig:Bethe}]{%
  \includegraphics[width=\columnwidth]{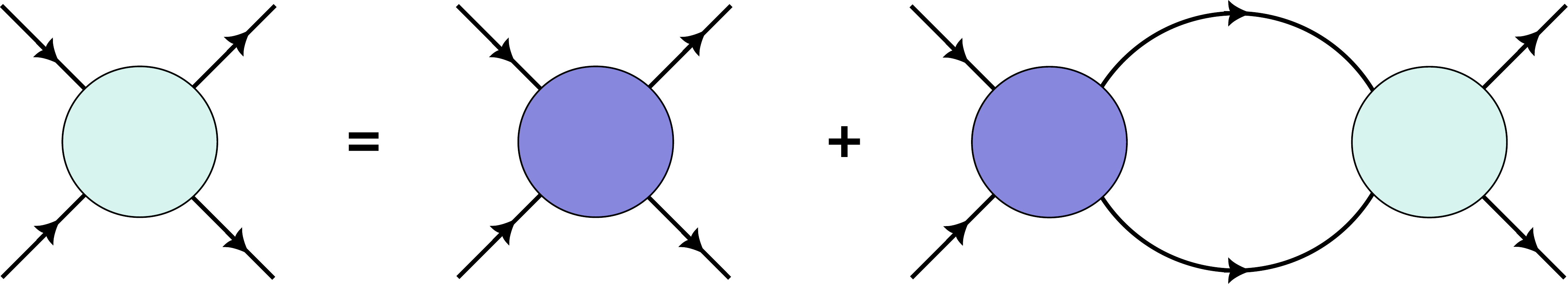}%
}
\hfill
\caption{ Diagrammatic  representations of the key processes contributing to the Boltzman equation. (a) Density response. The dashed (solid) arrow indicates sum over internal frequency (frequency, momenta, and spin). The smaller/red (larger/cyan) filled disk represents the external field, $\lambda$, that elicits the density response (total 4-fermion vertex, $\Gamma$). 
(b) The Bethe-Salpeter equation for $\Gamma$ in terms of the irreducible vertex.
} \label{Fig:BSE}
\end{figure}

To see how the external field modifies the local electronic density to linear order, we write a semi-classical total density as $n_{\sigma}(k,\bs r, t) = n_\sigma^0(\bs k) + \delta n_{\sigma}(k, \bs q_0)e^{i(\bs q_0 \cdot \bs r - \omega_0 t)} + \mbox{h.c.}$. 
Here $k \equiv (\bs k, \Omega)$ is the average momentum and energy of the incoming particles. 
We have further defined variation of the density as the expectation value, $\delta n_{\sigma}(\bs k, q_0) = \langle \psi (q_0)| \alpha^{\dagger}_{\bs k- \frac{1}{2} \bs q_0 \sigma} \alpha_{\bs k + \frac{1}{2} \bs q_0 \sigma}|\psi(q_0) \rangle$,  where $\alpha_{\bs k \sigma}$ is the quasiparticle operator, and $| \psi(q_0) \rangle = |\psi_0 \rangle + |\delta \psi (q_0) \rangle$ is the ground state $|\psi_0\rangle$ corrected by $|\delta \psi \rangle$ due to the external field. The density response, $\delta n_{\sigma}(\bs k, q_0) = \frac{1}{2\pi i}\int d\Omega \chi_{\sigma}(k, q_0)$, can be evaluated using the  diagram in Fig.\,\ref{Fig:BSE}(a), and it is given by 
\beq 
\chi_{\sigma}(k, q_0) = z_{\bs k}^{-1}  \Lambda_{\sigma}(k, q_0) G\left(k^-\right) G\left(k^+\right).
\eeq
Here $z_{\bs k}$ is the quasiparticle residue and is obtained by converting the quasiparticle operators in terms of the physical electron operators. $G(k^{\pm})\propto z_{\bs k}$ are the interacting propagators for shifted wave vector and energies $ k^{\pm} \equiv  (\bs k \pm \frac{1}{2} \bs q_0, \Omega \pm \frac{1}{2} \omega)$.
Scatterings among electrons renormalize $\lambda_\sigma$ [red disk in Fig.\,\ref{Fig:BSE}(a)] to  $\Lambda_{\sigma}$,  which is given by 
\begin{align}
& \Lambda_{\sigma}(k, q_0) = \lambda_\sigma(\bs k) \nonumber \\ 
&\qquad + \sum_{\bs k'\sigma' \Omega'}\lambda_{\sigma'}(\bs k') \Gamma_{\sigma,\sigma'}(k, k', q_0)  G\left(k^-\right) G\left(k^+\right),
\label{Lambda}
\end{align}
where $\Gamma_{\sigma, \sigma'}$ is the fully dressed 4-fermion vertex. 
In order to include renormalizations from both electron-electron and electron-impurity scatterings, we utilize the  Bethe-Salpeter equation [see Fig.\,\ref{Fig:BSE}(b)] to obtain $\Gamma_{\sigma,\sigma'}(k, k', q_0) = \hat{\Gamma}_{\sigma,\sigma'}(k, k', q_0) + \Delta \Gamma_{\sigma,\sigma'}(k, k', q_0) $  where 
\begin{align}
& \Delta \Gamma_{\sigma,\sigma'}(k, k', q_0) = \sum_{\substack{\bs k'' \sigma''\\\Omega''}}\hat{\Gamma}_{\sigma,\sigma''}(k,   k'', q_0) \nonumber \\ 
&  \times G\left(k''^-\right)  G\left(k''^+\right) \Gamma_{\sigma'',\sigma'}(k'', k', q_0)
\end{align}
and $\hat{\Gamma}_{\sigma,\sigma'}(k, k', q_0) $ (Fig.\,\ref{Fig:Spectral}) is the sum of all irreducible vertex diagrams including Coulomb and impurity scatterings. 

Eliminating the vertex part $\Lambda_{\sigma}(k, q_0)$ [Eq.\,(\ref{Lambda})] between the density response [Fig.\,\ref{Fig:BSE}(a)] and the Bethe-Salpeter equation [Fig.\,\ref{Fig:BSE}(b)] in favor of the irreducible vertex $\hat{\Gamma}_{\sigma,\sigma'}(k, k', q_0) $, we obtain the
Boltzmann transport equation for the quasiparticles,
\begin{align}
    dn_{\bm{p}}(\bm{x},t)/dt+I(n_{\bm{p}})=0 \label{B0}
\end{align}
with $d/dt=\partial_{t}+\dot{\bm{x}}\cdot\partial_{\bm{x}}\dot{\bm{p}}\cdot\partial_{\bm{p}}$, where $
    \dot{\bm{x}}=\partial \epsilon_{p}/\partial \bm{p}=\bm{v},
    \dot{\bm{p}}=-\partial\epsilon_{p}/\partial\bm{x}=e\bm{E}-\sum_{\bm{p}'}f_{\bm{p},\bm{p}'}\partial_{\bm{x}}\delta n_{\bm{p}'}
$, and
\begin{align}
\epsilon_{\bm{p}}(x)=\epsilon_{\bm{p}}^{0}-e\bm{E}\cdot\bm{x}+\sum_{\bm{p}'}f_{\bm{p},\bm{p}'}\delta n_{\bm{p}'}\label{energy}
\end{align}
denotes the total quasiparticle energy. $\epsilon^{0}_{\bm{p}}$ 
corresponds to the {non-interacting fermion} energy at global equilibrium. 
$\bm{v}$ denotes the quasiparticle velocity, $\bm{E}$ refers to the static electric field acting on the quasiparticles. 
$I(n_{\bm{p}})=I_{im}+I_{ee}$  consists of electron-impurity and electron-electron collision integrals. Note that the vertex parts can be solved exactly in the static long-wavelength limit of $\bs q_0, \omega_0 \rightarrow 0$ with constant $|\bs q_0|/\omega_0$~\cite{Nozieres1966}. Since the Ward identities constrain $\Lambda_{\bs k} \propto z_{\bs k}^{-1}$, it is clear from the expression of $\chi_{\sigma}(\bs k, \Omega, \bs q_0, \omega_0)$ that the density response, and as a consequence the Boltzmann equation, are independent of the quasiparticle residue. 

While the Boltzmann transport equation is useful to calculate the nonequilibrium average electronic behavior, it is insufficient to describe their fluctuations. To do this, we introduce a Langevin source term $\delta J^{ext}$ to the Boltzmann equation\cite{kogan1969}, which allows the room for quasiparticle fluctuations $n_{\bm{p}}\rightarrow n_{\bm{p}}+\delta n^{fl}_{\bm{p}}$ to give the Boltzmann-Langevin equation for a strongly correlated Fermi liquid,
\begin{align}
&{(\partial_{t}+\bm{v}\cdot\partial_{\bm{x}}+e\bm{E}\cdot\partial_{\bm{p}})\delta n_{p}^{fl}(x,t)+\delta I}=\label{BL}
\\&{-e\delta\bm{E}\cdot\bm{v}\partial_{\epsilon_{p}}n_{p}+\frac{\partial n_{p}} {\partial{{\epsilon_{p}}}}\bm{v}\cdot\sum_{p'} f_{p,p'}\partial_{\bm{x}}\delta n^{fl}_{p'}+\delta J^{ext}(p,x,t)}
\,\nonumber
\end{align}
where  {$\delta\bm{E}$ is the field fluctuation induced by quasiparticle fluctuations, and is determined self-consistetly through the Maxwell equation \cite{Nagaev1992}, $\nabla\cdot\delta{\bm{E}}= 4\pi\delta\rho^{fl}$, where $\delta\rho^{fl} =e \sum_{p}\delta{n}^{fl}_{p}$ is the charge fluctuation.} $\delta I$  represents the change of collision integral due to fluctuating quasiparticle distribution $\delta n_{p}^{fl}$. 

{We note that the Landau parameters appear explicitly in Eqs. (5), (6), and (7) through the kinetic terms,  and the renormalizations to the quasiparticle energy ($\epsilon_{\bs p}$). 
Additionally,  Eqs. (5) and (7) implicitly depend on the Landau parameters through the collision integrals ($I$), quasiparticle fluctuations ($\delta n_{p}^{fl}$) and field-fluctuations ($\delta E$). 
Here, $\delta \bm{E}$ is sensitive to the Landau parameters because it is generated by quasiparticle fluctuations $\delta n_{p}^{fl}$ that depend on the interactions among  the quasiparticles.
}

The fluctuations have zero mean value but have finite correlations. 
$\delta J^{ext}$ denotes the extraneous flux of particles in $\bm{p}$ state and equalsx
\begin{align}
    \delta J^{ext}(p,x,t)=\sum_{\bm{p}'}\delta J(p'p,x,t)-\delta J(pp',x,t).
\end{align}
It is the difference between flux from all $\bm{p}'$ states to $\bm{p}$ state and flux from $\bm{p}$ state to all $\bm{p}'$ states. We assume that the different fluxes are correlated when and only when the initial and final states are identical thereby following a Poisson  distribution of the form
\begin{align}
&\langle\delta J(p_{1}p_{1}',x_{1},t_{1})\delta J(p_{2}p_{2}',x_{2},t_{2})\rangle \nonumber\\
=&LA\delta_{p_1,p_2}\delta_{p_{1}',p_{2}'}\delta(x_{1}-x_{2})\delta(t_{1}-t_{2})J(p_{1}p_{1}',x_{1},t_{1}),
\end{align}
where $J(p_{1}p_{1}',x_{1},t_{1})$ is the mean flux of particles. {$L$ and $A$ are the length and cross section of the system} The presence of the Dirac delta functions in space-time reflect the fact that the duration and spatial extent of collisions is much smaller than the the electron-electron scattering lifetime and scattering length respectively.  We will use Eq.\,(\ref{BL}) to calculate the shot noise. 

\paragraph*{{\bf Steady state:}} Consider a diffusive correlated metallic wire with length $L$ and cross section $A$ $(L\gg\sqrt{A})$. We model the correlated metallic wire as a strongly interacting diffusive Fermi liquid with an applied voltage. An applied voltage drives the metal into a nonequilibrium steady state. The static electric potential energy $e\phi(x)=-e\bm{E}\cdot\bm{x}$ serves as an $s$ wave perturbation to the quasiparticle distribution at every position along the system.  The kinetics for the nonequilibrium system could be described by Boltzmann transport equation (\ref{B0}) for the quasiparticles.
The two ends of the wire stay in their own equilibrium:  {$n_{p}(\pm L/2,t)=f(\epsilon\pm eV/2, T_{bath}=0)$}, which serves as the boundary conditions to the Boltzmann equation, where $f$ is the Fermi-Dirac distribution function.

In the steady state, $\partial n_{\bm{p}}/\partial t=0$. In order to solve Eq.\,(\ref{B0}), we expand the Boltzmann equation in terms of $\delta \bar{n}(\bm{x},\bm{p})=n_{\bm{p}}(\bm{x})-n_{0}(\bm{x},\epsilon)$ in the regime where  $E_{F}\gg 1/\tau_{im}\gg 1/\tau_{ee}\gg D/L^{2}$. The first inequality corresponds to the condition for dilute impurities, and the following inequalities denote for the strong correlation regime. 
The distribution $n_{0}(\bm{x},\epsilon)$ stands for local equilibrium distribution and $\epsilon$ refers to true quasiparticle energy defined in Eq.\,(\ref{energy}). $\delta \bar{n}(\bm{x},\bm{p})$ corresponds to the departure from local equilibrium. This should be contrasted with the quasiparticle excitation $\delta n_{\bm{p}}(x)=n_{\bm{p}}(\bm{x})-n_{0}(\epsilon_{\bm{p}}^{0})$ in Eq.\,(\ref{energy}), where $n_{0}(\epsilon_{\bm{p}}^{0})$ refers to the {non-interacting fermion} distribution at global equilibrium without any position dependence. The two quantities are connected by $\delta \bar{n}(\bm{x},\bm{p})=\delta n_{\bm{p}}-\frac{\partial{n_{0}}}{\partial{\epsilon}}\sum_{\bm{p}'}f_{\bm{p},\bm{p}'}\delta n_{\bm{p}'}(\bm{x})$. Only the local equilibrium distribution $n_{0}(\bm{x},\epsilon)$ as a function of real quasiparticle energy $\epsilon$ could make the collision integral $I=I_{im}+I_{ee}$ vanish, and only $\delta \bar{n}(\bm{x},\bm{p})$, rather than $\delta{n}_{\bm{p}}(\bm{x})$, determines the current\cite{lifschitz1983,Nozieres1966}.  

{When quasiparticle scattering is strong such that electron-electron scattering length is small compared with system size: $l_{ee}\ll L$, it drives} the quasiparticles into local equilibrium, with a general position dependent Fermi-Dirac form $n_{0}(\epsilon,x)=f(\epsilon,T(x))$ such that $I(n_{0})=0$.
 Same as the equation in a Fermi gas~\cite{Rudin1995,Nagaev1995}, the explicit solution to {the Boltzmann equation} has the form 
\begin{align}
    T(x)=\frac{\sqrt{3}}{2\pi}eV\sqrt{1-\left(\frac{2x}{L}\right)^{2}}\label{tem}
\end{align}
with $V=EL$. The details of the derivation are shown in the `Supplementary Materials' (SM)~\cite{sm}. The local equilibrium distribution function $n_{0}=f(\epsilon,T(x))$ is plotted at zero and finite environment temperature in Fig.\,S1 in the SM~\cite{sm}.

\paragraph*{{\bf Shot noise:}} To calculate the shot noise, we model the mean particle flux as $J(p'p,x,t)=W(p'p)n_{p'}(x,t)(1-n_{p}(x,t))$, where $W(pp')$ is the scattering rate, which in the isotropic case  $W(pp')=\delta({\epsilon_{p}-\epsilon_{p'}})/(LAN_{F}\tau_{im})$.  The extraneous flux of quasiparticles could be connected with the quasiparticle fluctuation and thereby connects with the current fluctuation through Boltzmann-Langevin equation\,(\ref{BL}). We leave the details of the derivations to the SM~\cite{sm} and list the final results of the current noise,
\begin{align}
&S=2\int_{-\infty}^{\infty}dt\langle\delta I(t)\delta I(0)\rangle = \frac{\sqrt{3}}{2}GeV
\end{align}
where $T(x)$ is the local temperature in Eq.\,(\ref{tem}). $G=e^{2}DN_{F}A/L=ne^{2}\tau_{im}A/(m^{*}L)$ represents the conductance of the FL. 

We find that under electron correlation, the Landau parameters \textit{only renormalize the conductance}. Therefore the Fano factor $F=S/2eI$ is still $\sqrt{3}/4$ compared with the Fermi gas results in the hot electron regime, regardless of the interaction strength.  { This is despite the Landau parameter showing up in the quasiparticle energy, distribution function and its associated quasiparticle fluctuations (c.f. Eqs.(5, 6, \ref{BL}))}, {which is analogous to the robustness of the Weidemann-Franz law against interactions in a Fermi liquid
\cite{castellani1987}.} 

 \par
\paragraph*{{\bf Discussion:}} Several remarks are in order. First, in our calculations, we assume that the corrections to the quasiparticle density and energies are linear in the external perturbation (linear response).  The role of non-linearities in the response and the Fano factor dependence of the interaction parameters will be captured by higher order terms. Second, our calculations also assume intermediate mesoscopic length scales in accordance with experimental values. 
If the wire is too long, electron-phonon scattering must be considered, which acts to suppress the shot noise and Fano factor.
 If the length of the wire is much smaller than {electron-electron scattering length} $L_{ee}$, inelastic scattering does not contribute to the shot noise and Fano factor is automatically independent of the interaction strength and Landau parameters. 
Third, when retardation effects become important, as for example in the limit of lower electron densities, the Poissonian nature of the interaction fails. In this case, one must revisit the issue of the Fano factor dependence on the interaction parameters.   In  experiments on YbRh$_2$Si$_2$, the mesoscopic wire is quasi-three  dimensional where FL theory continues to hold. However, when the transverse dimensions 
of the wire are sufficiently reduced, FL theory eventually gives way to Luttinger liquid physics. In this case, the Fano factor generally depends on the dimensionless Luttinger liquid  parameters since the shot noise responds to an effective charge $ge$ with $g$ being the interaction parameter~\cite{Fisher1994, Buttiker2000}. {Finally, a large mean free path compared to the Fermi wavelength ($k_{F}l \sim 1000$;
see the SM)
indicates that disorder-related corrections are negligible.}

In Table I of the SM~\cite{sm} we have  contrasted our work with 
earlier works~\cite{Nagaev1992, Rudin1995, Nagaev1995, nagaev1998long} on shot noise in metals with quasiparticles. 
The contrast emphasizes that the combination of factors considered here helps identify
the shot noise Fano factor as a universal ratio, akin to
the Wiedemann-Franz law for Fermi liquids, especially in
the strong correlation regime.

To conclude, we have shown that the Fano factor of a strongly correlated Fermi liquid, defined by the ratio of its average current fluctuations to its average current, equals that of a Fermi gas in the linear response regime. More specifically, our results demonstrate that the Poissonian nature of the instantaneous Coulomb interaction and charge conservation dictate a Fano factor that is independent of the Landau parameters or (however small) quasiparticle residue respectively. This has important consequences to recent shot noise experiments in the heavy fermion material YbRh$_2$Si$_2$ where a strong suppression of the Fano factor was observed even when 
the effect of electron-phonon coupling is negligible~\cite{Natelson2022}. 
The existence of a quasiparticle interpretation is a sufficient requirement for our analysis to hold.  Thus, any suppression of the Fano factor below that of a correlated diffusive metal ($F=\frac{\sqrt{3}}{4}$) strongly suggests the loss of quasiparticles in YbRh$_2$Si$_2$.
More generally, our work points to shot noise Fano factor as a powerful characterization of strongly correlated metallic phases and materials.

{
\textit{Note added:} After this manuscript was submitted for publication, recent works became available that addressed the shot noise when electrons are coupled to collective bosons 
\cite{Niko2023,wu2023,Wang2024}.
}

\paragraph*{{\bf Acknowledgements: }}
This work was supported 
primarily by the National Science Foundation under Grant No.\ DMR-2220603 (Y.W.,S.S.),
by the Air Force Office of Scientific Research under Grant No.\ FA9550-21-1-0356 (C.S.),
and by the Robert A. Welch Foundation Grant No.\ C-1411 
and 
the Vannevar Bush Faculty Fellowship ONR-VB N00014-23-1-2870 (Q.S.). C.S. acknowledges support from Iowa State University and Ames National Laboratory start up funds. 
L.C. and D. N. acknowledge support by
US Department of Energy, Basic Energy Sciences, under award No. DE-FG02-06ER46337.
S.P. acknowledges funding from the Austrian Science Fund (projects No.\ I4047, 29279 and the Research unit QUAST-FOR5249) and 
the European  Microkelvin  Platform  (H2020 project No.\ 824109).
Q.S. acknowledges the hospitality of the Aspen Center for Physics,
which is supported by NSF grant No. PHY-2210452.


\clearpage
\appendix
\onecolumngrid
\section{Supplementary Material}
\subsection{Additional details for the derivation of the Boltzmann  equation}
As noted in the main text, the Boltzmann equation follows from eliminating the dressed external vertex, $\Lambda$, from the explicit  expression of density fluctuation, $\delta n_\sigma$.
Since we want to describe the dynamics of fluctuations that live on the Fermi surface, we seek a solution of the form~\cite{Nozieres1966}, 
\begin{align}
\delta n_\sigma(\mathbf k, q_0) = \nu(\hat k, q_0) \delta(\epsilon_{\mathbf k} - \mu),
\end{align}
where $\hat k$ represents Fermi surface coordinates.
This leads to 
\begin{align}
\delta n_\sigma(\mathbf k, q_0) = z_{\mathbf k} \delta(\epsilon_{\mathbf k} - \mu) \left[ \frac{\Lambda^1_{\sigma}(\hat k, q_0) + \Lambda^2_{\sigma}(\hat k, q_0)}{2i \Gamma_{\text{imp}} + \omega_0 - \mathbf v_{k} \cdot \mathbf q_0} \omega_0 - \Lambda^0_{\sigma}(\hat k, q_0)
\right],
\label{eq:dl-n}
\end{align}
where $\Gamma_{\text{imp}}$ is the rate of scattering between the electrons and the impurities, $\Lambda_\sigma^0(\hat k)$ is the solution for $\Lambda_\sigma(\hat k, q_0)$ at $\omega_0 = 0$, and   $\Lambda^{1,2}_\sigma$ are obtained by solving the Bethe-Salpeter equation for  $\Lambda_\sigma$, which takes the general form
\begin{align}
\Lambda_\sigma(k, q_0) = \Lambda_\sigma^1(k, q_0) + \Lambda_\sigma^2(k, q_0) \Theta(\omega_0 - 2 |\Omega - \mu|).
\end{align}
On the Fermi surface a recursive relationship for $\Lambda^n$'s is obtained~\cite{Nozieres1966}
\begin{align}
& \Lambda_\sigma^1(\hat k, q_0) = \Lambda_\sigma^0(\hat k) + \frac{\omega_0}{8\pi} \sum_{\sigma'} \int d\hat{k}' \frac{X_{\sigma, \sigma'}(\hat k, \hat k')}{2i\Gamma_{\text{imp}} + \omega_0 - \mathbf v_{k'} \cdot \mathbf q_0}  \left[ \Lambda_{\sigma'}^1(\hat k', q_0) + \Lambda_{\sigma'}^2(\hat k', q_0)\right] \\
& \Lambda_\sigma^2(\hat k, q_0) =  \frac{i}{2\pi}  \int d\hat{k}' \frac{Y(\hat k, \hat k')}{2i\Gamma_{\text{imp}} + \omega_0 - \mathbf v_{k'} \cdot \mathbf q_0}  \left[ \Lambda_{\sigma}^1(\hat k', q_0) + \Lambda_{\sigma}^2(\hat k', q_0)\right].
\end{align}
Here $X_{\sigma, \sigma'}(\hat k, \hat k')$ is a function of Fermi surface coordinates and fully determined by Landau parameters. 
By contrast, $Y(\hat k, \hat k')$ encodes contributions that exist only due to   electron-impurity scatterings. 
By adding $\Lambda^1_\sigma$ and $\Lambda^2_\sigma$, we replace their sum  in 
Eq.\,(\ref{eq:dl-n}) to obtain the Boltzmann equation in frequency-momentum space with the collision integral, 
\begin{align}
I(n_{\mathbf k}) = 2i \int \frac{d\hat{k}'}{4\pi} Y(\hat k, \hat k') \delta\bar{n}(\mathbf k')  - 2i \Gamma_{\text{imp}} \delta\bar{n}(\mathbf k),
\end{align}
where $\bar{n}(\mathbf k)$ is defined in the main text. 
In Eq. (4) we express the Boltzmann equation in real-time and coordinate space. 

We note that the contribution to $I$ above entirely results from electron-impurity collisions.
In our calculations we model it as
\begin{align}
I_{im} = \sum_{p'}W(pp')[n_{p}(x,t)(1-n_{p'}(x,t))-n_{p'}(x,t)(1-n_{p}(x,t))], \label{eq:Iimp}
\end{align}
where $W$ refers to the scattering rate for electron-impurity collisions, and propotional to $Y$.
In principle, electron-electron collisions also contribute to $I$, but they vanish on the Fermi surface at $T_{\text{bath}}=0$.
An applied voltage, however, generates scatterings among electrons at $T_{\text{bath}}=0$, which results in a finite local temperature $T(x)$ [see Sec. II].
Consequently, $I$ acquires a contribution resulting from the applied voltage,
\begin{align}
I_{ee} = \sum_{2,3,4}\widetilde{W}
(12;34)\delta_{\bm{p}+\bm{p}_{2},\bm{p}_{3}+\bm{p}_{4}}\delta{(\epsilon+\epsilon_{2}-\epsilon_{3}-\epsilon_{4})}[n_{3}n_{4}(1-n_{1})(1-n_{2})-n_{1}n_{2}(1-n_{3})(1-n_{4})],
\label{eq:Iee}
\end{align}
where $\widetilde{W}
$ refers to the transmission probability for electron-electron collisions.
Thus, the net $I$ in the presence of an applied voltage is $I = I_{im} + I_{ee}$.

\subsection{Local Equilibrium Solution of the Boltzmann Equation}
In this subsection we solve the Boltzmann equation in the strong quasiparticle scattering regime. When $I_{im}\gg I_{ee}\gg dn/dt$, The zeroth order solution of the Boltzmann equation:
\begin{align}
    n_{0}(\epsilon-\mu(x), x)=\frac{1}{1+e^{\frac{\epsilon-\mu(x)}{T(x)}}}
\end{align}
is determined by
\begin{align}
    I_{im}+I_{ee}=0
\end{align}
where $I_{im}$ and $I_{ee}$ refer to impurity-quasiparticle and quasiparticle-quasiparticle scattering in Eqs.(\ref{eq:Iimp}, \ref{eq:Iee}). $\mu(x)$ and $T(x)$ are unknown position dependent functions. Next we expand the Boltzmann equation  to first order and second order in O($n-n_{0})$.
After the expansion, we get two equations in terms of $n_{0}$ and $\delta \bar{n}$,
\begin{align}
&\bm{v}\cdot\partial_{\bm{x}}n_{0}(\bm{x},\epsilon_{\bm{p}})+e\bm{E}\cdot\bm{v}\partial_{\epsilon}n_{0}(\bm{x},\epsilon_{\bm{p}})=\frac{\delta \bar{n}}{\tau}\label{B1}\\
&\bm{v}\cdot\partial_{\bm{x}}\delta \bar{n}+e\bm{E}\cdot\partial_{\bm{p}}\delta \bar{n}=0\label{B2}
\end{align}
where $\epsilon_{\bm{p}}=\epsilon_{\bm{p}}^{0}+\sum_{\bm{p}'}f_{\bm{p},\bm{p}'}\delta n_{\bm{p}}$ is the quasiparticle energy which includes Landau parameters, and $\delta \bar{n}_{p}=\delta {n}_{p}-\frac{\partial n_{0}}{\partial{\epsilon}}\sum_{p'}f_{p,p'}\partial_{x}\delta n_{p'}$. In the case where the conductance is mainly contributed by impurity scattering, $\tau_{im}\ll\tau_{ee}$, we can  
relace $\tau$ with $\tau_{im}$ in Eq.\,({\ref{B1}}).
Then we follow the analysis of Kozub and Rudin \cite{Rudin1995}. Substituting Eq.\,(\ref{B1}) into (\ref{B2}), and integrating with ${\epsilon}d\epsilon$. 
We get the diffusion equation in terms of the local equilibrium temperature at leading order in $E$,
\begin{align}
    \frac{\pi^{2}}{6}\frac{\partial^{2}}{\partial x^{2}}T^{2}(x)+(eE)^{2}=0 \label{diff}
\end{align}
with boundary conditions $T(\pm L/2)=0$.
The solution to Eq.(\ref{diff}) is straitforward:
\begin{align}
    T(x)=\frac{\sqrt{3}}{2\pi}eV\sqrt{1-\left(\frac{2x}{L}\right)^{2}}.
\end{align}

\subsection{Fluctuations of the nonequilibrium Boltzmann-Langevin equation for a Fermi liquid}
As described in the main text, the fluctuations of the nonequilibrium quasiparticles are described by the Boltzmann-Langevin equation in Eq.\,(7) in the main text:
\begin{align}
&(\partial_{t}+\bm{v}\cdot\partial_{\bm{x}} +e\bm{E}\cdot\partial_{\bm{p}})\delta n_{p}^{fl}(x,t)+\delta I=-e\delta\bm{E}\cdot\bm{v}\partial_{\epsilon_{p}}n_{p}+\frac{\partial n_{p}}{\partial{{\epsilon}}}\bm{v}\cdot\sum_{p'} f_{p,p'}\partial_{\bm{x}}\delta n^{fl}_{p'}+\delta J^{ext}(p,x,t)
\,
\label{BL}\end{align}
where $\delta J^{ext}$ denotes the extraneous flux of the particles in $\bm{p}$ state, and $\delta I$ represents the change of the collision integral due to fluctuating quasiparticles. $\delta\bm{E}$ is the field fluctuation induced by quasiparticle fluctuations, and is determined self-consistetly~\cite{Nagaev1992}, $\nabla\cdot\delta{\bm{E}}= 4\pi\delta\rho^{fl}$, where $\delta\rho^{fl} =e \sum_{p}\delta{n}^{fl}_{p}$ is the charge fluctuation. These fluctuations have finite correlations with zero mean values, $\langle\delta{n}^{fl}\rangle=\langle\delta I\rangle=\langle\delta J^{ext}\rangle=0$. 
Since the current is mainly contributed by the elastic collision $(\tau_{im}\ll\tau_{ee})$, we only consider 
flux from the elastic collisions,
\begin{align}
     I_{imp}=&\sum_{p'}J(p'p)-J(pp')\\
    J(p'p,x,t)=&W(p'p)n_{p'}(x,t)(1-n_{p}(x,t))\label{mf}
\end{align}
where $W(pp')$ is the scattering rate. In the isotropic case, it is given as follows:
\begin{align}
    W(pp')=\frac{\delta(\epsilon_{p}-\epsilon_{p'})}{LAN_{F}\tau_{im}}.
\end{align}

The correlation of $\delta J^{ext}$ is determined only by the fact that each electron scattering event is independent of one another~\cite{kogan1969}.
Thus, each scattering is only self-correlated.
The flux into the $\bm{p}$ state could be expressed by the subtraction between the flux from all $\bm{p}'$ states to the $\bm{p}$ state and the flux from the $\bm{p}$ state to all $\bm{p}'$ states,
\begin{align}
    \delta J^{ext}(p,x,t)=\sum_{\bm{p}'}\delta J(p'p,x,t)-\delta J(pp',x,t) \, .
\end{align}
Different fluxes are correlated when and only when the initial and final states are identical, following Poissonian statistics:
\begin{align}
\langle\delta &J(p_{1}p_{1}',x_{1},t_{1})\delta J(p_{2}p_{2}',x_{2},t_{2})\rangle=LA\delta_{p_1,p_2}\delta_{p_{1}',p_{2}'}\delta(x_{1}-x_{2})\delta(t_{1}-t_{2})J(p_{1}p_{1}',x_{1},t_{1}) \, .
\end{align}
Then 
\begin{align}
    &\langle\delta J^{ext}(p_{1},x_{1},t_{1})\delta J^{ext}(p_{2},x_{2},t_{2})\rangle=LA\delta(x_{1}-x_{2})\delta(t_{1}-t_{2})\{\delta_{p_{1},p_{2}}\sum_{q}[J(p_{1}, q)+J(q,p_{1})]-J(p_{1},p_{2})-J(p_{2},p_{1})\}
\label{cor}\end{align}
where $J(p_{1},p_{2})$ denotes the mean quasiparticle flux defined in Eq.\,(\ref{mf}).
The Boltzmann-Langevin equation\,(\ref{BL}) could be re-expressed as
\begin{align}
    &\partial_{t}\delta n^{fl}+\bm{v}\cdot\partial_{\bm{x}}\delta \bar{n}^{fl}+e\bm{E}\cdot\partial_{\bm{p}}\delta n_{p}^{fl}(x,t)+\delta I=-e\delta\bm{E}\cdot\bm{v}\partial_{\epsilon_{p}}n_{0}+\delta J^{ext}\label{BL2},
\end{align}
where $\delta \bar{n}_{p}^{fl}=\delta {n}_{p}^{fl}-\frac{\partial n_{0}}{\partial{\epsilon}}\sum_{p'}f_{p,p'}\partial_{x}\delta n_{p'}^{fl}$, which is similar to the relation between $\delta n$ and $\delta \bar{n}$ in the Landau transport equation of 
a
FL~\cite{lifschitz1983,Nozieres1966}. Following the analysis of Nagaev~\cite{Nagaev1992}, we express the fluctuation $\delta \bar{n}$ and the source as the sum of symmetric and antisymmetric components in momentum space,
\begin{align}
    \delta \bar{n}^{fl}(\bm{p},\bm{x})=&\delta \bar{n}_{0}^{fl}(\epsilon,\bm{x})+\hat{\bm{n}}\cdot\delta\bar{\bm{n}}_{1}^{fl}(\epsilon,\bm{x})\\
    \delta J^{ext}(\bm{p},\bm{x})=&\delta j^{ext}(\epsilon,\bm{x})+\hat{\bm{n}}\cdot\delta\bm{J}(\epsilon,\bm{x})\label{j}
\end{align}
where $\epsilon$ is the real quasiparticle energy, and $\hat{\bm{n}}$ refers to the unit vector in momentum space. Note that when transforming from $(\bm{p}, \bm{x})$ to $(\epsilon, \bm{x})$ coordinates, the electric field $\bm{E}$ and $\delta\bm{E}$ related terms are absorbed into quasiparticle energy $\epsilon$. Similar to what happens with the steady state current in the FL~\cite{lifschitz1983,Nozieres1966}, the fluctuation of the current density $\delta\bm{j}^{fl}(\bm{x},t)$ is only determined by $\delta \bar{\bm{n}}_{1}^{fl}$, 
\begin{align}    
\delta\bm{j}^{fl}=&e\sum_{ps}\bm{v}\delta \bar{n}^{fl}=e\sum_{p}\bm{v}(\hat{\bm{n}}\cdot \delta \bar{\bm{n}}_{1}^{fl}).\label{cufl}
\end{align} 
Integrating Eq.\,(\ref{BL2}) with  $\hat{\bm{n}}d\hat{\bm{n}}$ gives
\begin{align}
    \delta\bm{J}=-\frac{\delta \bar{\bm{n}}^{fl}_{1}}{\tau_{im}}+v\partial_{\bm{x}}\delta{\bar{n}_{0}^{fl}}\label{flux}
\end{align}
where we have used the relaxation time approximation for the impurity collision integral, $\delta I_{im}=-\hat{\bm{n}}\cdot\delta \bar{\bm{n}}^{fl}_{1}/\tau_{im}$. Combining the current formula in Eq.\,(\ref{cufl}) with the flux in Eq.\,(\ref{flux}),
\begin{align}
    \delta I(t)=&\frac{1}{L}\int d^{3}x\delta j(x,t)=\frac{ev_{F}N_{F}\tau_{im}}{3L}\int d^{3}xd\epsilon\delta J(\epsilon,x,t)\, . \label{ifluc}
\end{align}
In general, the fluctuating quasiparticles $\delta n^{fl}$ could generate fluctuating electric fields $\delta E$, thereby contributing additional current. However, only the
current generated by the Langevin source in Eq.\,(\ref{ifluc}) contributes to the noise \cite{kogan2008}.

\begin{figure}[!t]
\centering

\subfloat[\label{fig:hot}]{%
  \includegraphics[width=0.45\columnwidth]{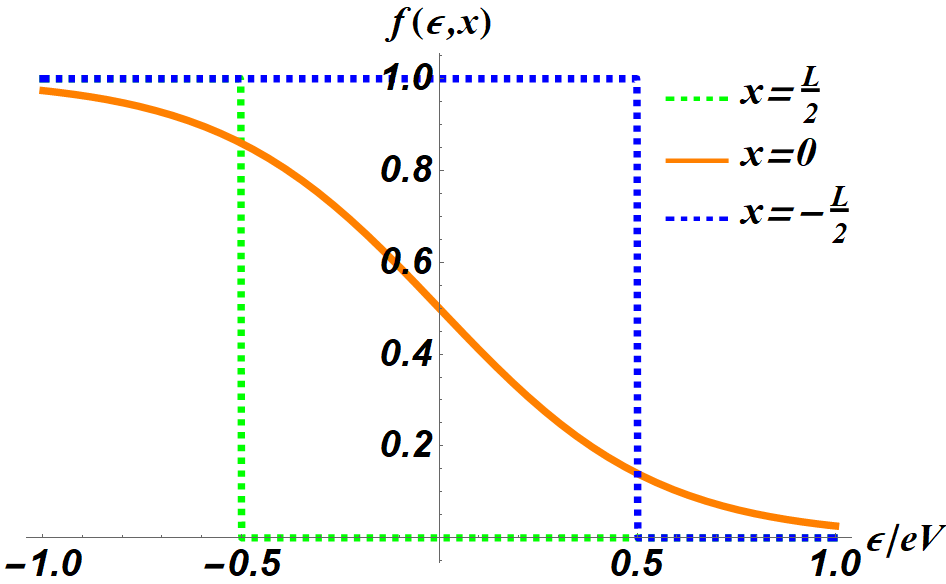}%
}
\hfill
\subfloat[\label{fig:hot}]{%
  \includegraphics[width=0.45\columnwidth]{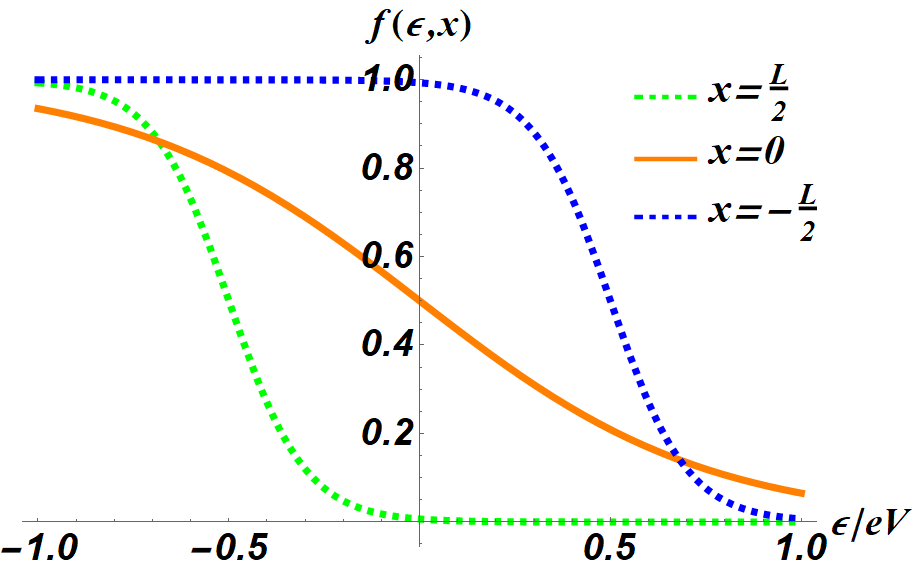}%
}
\caption{Plot of the local equilibrium distribution function with kinetic energy in the correlated metal with (a) $ T_{bath}=0$, (b) $T_{bath}>0$ at $x=L/2$ (Green), $x=0$ (Orange) and $x=-L/2$ (Blue) respectively.} \label{Fig:distribution}
\end{figure}

On the other hand, the correlation of $\delta\bm{J}$ is determined by substituting Eq.\,(\ref{j}) into Eq.\,(\ref{cor}), and integrating Eq.\,(\ref{cor}) with $\hat{\bm{n}}d\hat{\bm{n}}\hat{\bm{n}}'d\hat{\bm{n}}'$ with the relation $\delta_{pp'}=\delta(\hat{\bm{n}}-\hat{\bm{n}}')\delta(\epsilon-\epsilon')/N_{F}$
\begin{align}
    &\langle\delta J_{\alpha}(\epsilon,\bm{x},t)J_{\beta}(\epsilon',\bm{x}',t')\rangle=\delta_{\alpha\beta}\delta(t-t')\delta(\bm{x}-\bm{x}')\delta(\epsilon-\epsilon')\frac{6}{\tau_{im}N_{F}}n(\epsilon,\bm{x})[1-n(\epsilon,\bm{x})],\label{cor2}
\end{align}
where the prefactor $6$ comes from the first two terms of Eq.\,(\ref{cor}) and the relation $\hat{n}_{\alpha}^{2}=1/3$.
The current noise is then calculated by combining Eqs.\,(\ref{ifluc},\ref{cor2}), 
\begin{align}
    S=&2\int_{-\infty}^{\infty}dt\langle\delta I(t)\delta I(0)\rangle\\
    =&\frac{4G}{L}\int_{-L/2}^{L/2} dx d\epsilon n(x,\epsilon)[1-n(x,\epsilon)]\\
    =&\frac{4G}{L}\int_{-L/2}^{L/2} dx T(x)
\end{align}
where $T(x)=\int d\epsilon n(1-n)$ represents the local equilibrium temperature, $G=\frac{e^{2}v_{F}^{2}\tau_{imp}N_{F}A}{3L}=\sigma A/L$ marks the conductance of a FL, and $\sigma=\frac{ne\tau_{im}}{m^{*}}$ denotes the conductivity. 
We note that, due to the local equilibraton, the distribution function develops a position
dependence, as shown in Fig.\,\ref{Fig:distribution}. 

If the system is in a global equilibrium, the current noise equals the thermal noise $S=4GT_{bath}$. In the strongly correlated nonequilibrium FL with a nonzero external voltage bias at $T_{bath}=0$, as shown in the main text, the current noise is the shot noise $S=\sqrt{3}GeV/2$, with Fano factor 
\begin{align}
    F=\frac{S}{2eI}=\frac{\sqrt{3}}{4}.
\end{align}

\subsection{Comparison with previous literature}
Here we summarize the comparison of our work with prior literature on
the Fano factors and associated physical processes through Tab.~\ref{tab:my_label}.

\begin{table}[!h]
    \centering
    \begin{tabular}{ccc}
    \textbf{Fano factor} & \textbf{Physical Processes} & \textbf{References} \\
        $\frac{1}{3}$ & e-imp scattering in $I(n_p)$ & \cite{Nagaev1992} \\[2em]
         $\frac{\sqrt 3}{4}$& e-imp + e-e scattering in $I(n_p)$ &  \cite{Rudin1995},\cite{Nagaev1995}\\[2em] 
         $\frac{1}{3}$ & $\substack{\mbox{e-imp in $I(n_p)$ + electron gas } \\
\mbox{+  static screening of the electric field} \\ \mbox{by the long range Coulomb interaction }}$ & \cite{nagaev1998long}\\ [2em]
         $\frac{\sqrt 3}{4}$ & $\substack{\mbox{ e-imp + e-e scattering in $I(n_p)$} \\ \mbox{Derived kinetic term in BLE } \\ \mbox{with q.p. weight and Landau parmeters} \\
         \mbox{+ Landau parameters in fluctuations $\delta E$, $\delta n_p^{fl}$ } \\
         \mbox{derived from noise term $\delta J_{ext}(x, t)$ in Eq.~\ref{BL}} }$& Our work \\ [2em]
    \end{tabular}
    \caption{Comparison of our work 
    with prior literature. Here ``BLE" stands for Boltzmann-Langevin equation. }
    \label{tab:my_label}
\end{table}

\subsection{Estimation of the normalized mean free path $k_{\rm F} \ell$ in YbRh$_2$Si$_2$}

YbRh$_2$Si$_2$ is a stoichiometric material with a very 
small residual resistivity $\rho_0$,
which is on the order of 
$1\,\mu$ Ohm-cm (reaching as low as $0.55\,\mu$ Ohm-cm;
see Fig 4b, Ref~\cite{gegenwart2008quantum}). 
From $\rho_0$, we use the standard procedure 
(eg., Ref~\cite{abrahams2011quantum}, see p.2)
to estimate the normalized mean free path $k_{\rm F} \ell$.

The system being tetragonal, we can estimate the sheet resistance $R_{\rm sheet}$  from dividing $\rho_0$ by the spacing between the Yb-containing layers $c$. This separation is half
of the 
inter-layer lattice constant,
%
$9.86$ \AA.
The procedure leads to 
\begin{align}
R_{\rm sheet} =
 \frac{\rho_0 } {c} \approx 20 \, {\rm Ohm}\, .
\label{R-sheet}
\end{align}

Dividing 
the sheet resistance by the quantum resistance $h/e^2 \approx 26,000$ Ohm 
yields the following estimate for $1/ k_F \ell$
\begin{eqnarray}
\frac{1}{k_F \ell}
= \frac{R_{\rm sheet} } {h/e^2} 
\approx \frac{20}{2.6 \times 10^4} \approx 0.77 \times 10^{-3} \, .
\label{1overkFl}
\end{eqnarray}
This value 
suggests that the disorder in
YbRh$_2$Si$_2$ is very weak.

\bibliography{Noise.bib}

 \end{document}